\documentclass{jfm}
\usepackage{natbib}
\usepackage{graphicx}
\usepackage{subfigure}
\usepackage{bm}

\ifCUPmtlplainloaded \else
  \checkfont{eurm10}
  \iffontfound
    \IfFileExists{upmath.sty}
      {\typeout{^^JFound AMS Euler Roman fonts on the system,
                   using the 'upmath' package.^^J}%
       \usepackage{upmath}}
      {\typeout{^^JFound AMS Euler Roman fonts on the system, but you
                   dont seem to have the}%
       \typeout{'upmath' package installed. JFM.cls can take advantage
                 of these fonts,^^Jif you use 'upmath' package.^^J}%
      }
  \else
  \fi
\fi

\ifCUPmtlplainloaded \else
  \checkfont{msam10}
  \iffontfound
    \IfFileExists{amssymb.sty}
      {\typeout{^^JFound AMS Symbol fonts on the system, using the
                'amssymb' package.^^J}%
       \usepackage{amssymb}%
         
         \let\geq=\geqslant
      }{}
  \fi
\fi

\ifCUPmtlplainloaded \else
  \IfFileExists{amsbsy.sty}
    {\typeout{^^JFound the 'amsbsy' package on the system, using it.^^J}%
     \usepackage{amsbsy}}
    {}
\fi

\newcommand{\rem}[1]{}

\newsavebox{\astrutbox}
\sbox{\astrutbox}{\rule[-5pt]{0pt}{20pt}}

\newcommand{\pai}{\partial_i}
\newcommand{\paj}{\partial_j}
\newcommand{\pak}{\partial_k}
\newcommand{\grad}{\mathbf \nabla}
\newcommand{\paip}{\partial_{i'}}
\newcommand{\pajp}{\partial_{j'}}
\newcommand{\pakp}{\partial_{k'}}
\newcommand{\parj}{\partial^2_{r_j}}

\newcommand{\commentout}[1]{}

\def\ni{\noindent}
\def\bom{\bm{\omega}}

\title[The Two-point Correlation of Potential Vorticity]{On the Two-point Correlation of Potential Vorticity in Rotating and Stratified Turbulence} 

\author[S. Kurien, L.~Smith, B.~Wingate]{%
S\ls U\ls S\ls A\ls N\ns K\ls U\ls R\ls I\ls E\ls N$^1$,\ns%
L\ls E\ls S\ls L\ls I\ls E\ls \ns S\ls M\ls I\ls T\ls H$^2$ \and B\ls E\ls T\ls H\ns W\ls I\ls N\ls G\ls A\ls T\ls E$^3$ }

\affiliation{$^1$ Theoretical Division, Los Alamos National Laboratory,
Los Alamos, NM 87545, USA \\[\affilskip] 
$^2$ Departments of Mathematics and Engineering Physics, University of Wisconsin, Madison, WI 53706, USA \\[\affilskip]
$^3$ Computer and Computational Sciences Division, Los
Alamos National Laboratory, Los Alamos, NM 87545, USA \\[\affilskip]
}

\date{\today}

\begin{document}

\maketitle

\begin{abstract}
A framework is developed to describe the
two-point statistics of potential vorticity in rotating and stratified
turbulence as described by the Boussinesq equations. The
K\'arm\'an-Howarth equation for the dynamics of the two-point
correlation function of potential vorticity reveals the possibility of
inertial-range dynamics in certain regimes in the Rossby, Froude,
Prandtl and Reynolds number parameters.  For the case of large Rossby
and Froude numbers, and for the case of quasi-geostrophic dynamics, a
linear scaling law with 2/3 prefactor 
is derived for the third-order mixed correlation
between potential vorticity and velocity, a result that is analogous to the
Kolmogorov 4/5-law for the third-order velocity structure
function in turbulence theory. 


\end{abstract}

\section{Introduction}
Potential vorticity $q = \bm{\omega}_a \cdot {\nabla} \rho$, where
$\bm{\omega}_a$ is the total vorticity and $\rho$ is the density, is a
lagrangian invariant in rotating and stratified flows
(\cite{Ertel42}).   In the quasi-geostrophic (QG) limit, in which the
rotation and stratification effects are strong, the potential
vorticity evolution entirely determines the dynamics of quantities
such as wind speed, pressure and temperature fields (see for example,
\cite{HOSKINSBJ:Onuas, RHINESPB:VORDOT, 
MullerP:Ertpvt, HAYNESPH:Oncai}).
Instead of studying $q$ as a local invariant in the traditional way,
we here focus on the  global statistical description of the potential
vorticity field in the style of the Kolmogorov 1941 theory 
for the
statistical description of the velocity field (\cite{K41}).
\commentout{The use of potential vorticity to reveal
fundamental aspects of turbulence from the statistical hydrodynamics
perspective has, however, been somewhat sparse. There are limitations
to using $q$ itself as the statistical variable} We treat
potential vorticity as a statistical variable and consider the
dynamics of its second-order moment in the triply periodic
or infinite domain. Thus the
potential enstrophy $Q = \langle q^2 \rangle /2$ is the conserved quantity of interest, 
where $\langle\cdot\rangle$ denotes a suitable average (over ensembles
or over the flow domain).

\cite{Charney71}
showed analytically that in the QG limit, conservation of energy
and potential enstrophy implies an inverse cascade of
energy with energy spectrum scaling as $E(k)\sim k^{-5/3}$ for the low
wavenumbers, and a forward cascade of potential enstrophy with $E(k)
\sim k^{-3}$ at high wavenumbers. The strong constraint on the
energy by $Q$ is analogous to the constraint on energy by enstrophy in
2D turbulence, resulting in the well-known inverse cascade of energy,
the forward cascade of enstrophy, and the corresponding scalings of
the energy spectrum (\cite{Kraichnan71b}). \cite{HerKerRot94}
performed moderate-resolution, spectral simulations of
non-rotating, non-stratified turbulent flow with a passive scalar
$\theta$, in which the Ertel potential vorticity $q =
\bm{\omega} \cdot \grad \theta$ with $\bm{\omega} = \grad \times \bm{u}$. 
They studied the evolution of the
potential enstrophy spectrum and demonstrated that the transfer terms
and the dissipation terms are not separated in wavenumber
space. They argued therefore, that an `inertial' range of scales
dominated by the flux of potential enstrophy does not seem
possible. The concept of an inertial range is a cornerstone of the K41
theory of turbulence. Therefore it seemed futile, in a
certain sense, to further explore statistical approaches for
potential vorticity dynamics.

To revisit the possibility of potential enstrophy intertial-range
dynamics, we 
depart from previous approaches in two ways.  First, we confine
ourselves to statistics in $physical$ space,
and undertake a novel study of the evolution equations for the 
two-point spatial correlation function of potential vorticity. 
Second, we examine, in addition to the QG and non-rotating, non-stratified
limits, four other regimes of interest in the rotation and
stratification parameters.  Our approach has revealed that some of
these regimes $do$ allow for the existence of an inertial range of
potential enstrophy.  Furthermore, when we take a prescribed sequence of limits
in the non-dimensional parameters and assume local isotropy, then
exact scaling laws for small-scale potential vorticity statistics
emerge.  These new laws are analogous to existing laws that form the
backbone of statistical hydrodynamics in 3D non-rotating, non-stratified
flows.

Following K41 turbulence theory, the \cite{KH38}
equations for the two-point correlation functions of velocity lead to
the concept of an `inertial range' of scales dominated by net
downscale flux of kinetic energy by nonlinear transfer with little
dissipative loss.  The inertial-range concept was
extended to the flux of passive scalar variance by
\cite{Yaglom49}, and more recently to helicity flux by
\cite{Chkhetiani96} (see also
\cite{LPP97, GPP00, Kurien03}). Under the assumptions of statistical 
homogeneity and statistical isotropy of the small scales (local isotropy), all these results take the form of exact scaling
laws for the appropriate two-point third-order correlations as
follows:
\begin{subeqnarray}
\langle (u_L(\bm{x}+\bm{r}) - u_L(\bm{x}))^3\rangle &= -\displaystyle\frac{4}{5}\varepsilon r;&~\varepsilon = 2\nu \langle\grad\bm{u}\cdot \grad\bm{u}\rangle \\
\langle (u_L(\bm{x}+\bm{r}) - u_L(\bm{x}))(\theta(\bm{x}+\bm{r}) - \theta(\bm{x}))^2\rangle &= -\displaystyle\frac{4}{3}\varepsilon_\theta r;&~\varepsilon_\theta = 2\alpha \langle\grad\theta\cdot\grad\theta\rangle \\
\langle (u_L(\bm{x}+\bm{r}) - u_L(\bm{x}))(\bm{u}_T(\bm{x}+\bm{r}) 
\times \bm{u}_T(\bm{x}))\rangle &= \displaystyle\frac{2}{15}h r^2;&~ h = 2\nu \langle \grad\bm{u}\cdot \grad\bom\rangle 
\label{scalinglaws}
\end{subeqnarray}
where $\bm{r}$ is the separation vector between
the two measurement points; the velocity component along $\bm{r}$ is
called the longitudinal velocity $u_L = \bm{u} \cdot \hat{\bm{r}}$ (we
will also refer to its vector representation $\bm{u}_L = u_L \hat{\bm{r}}$);
the component orthogonal to $\bm{r}$ is called the transverse velocity
$\bm{u}_T = \bm{u} - u_L~\hat{\bm{r}}$; $\theta$ is the passive
scalar.  The quantities $\varepsilon$, $\varepsilon_\theta$ and $h$
are, respectively, the mean dissipation rates of kinetic energy,
passive scalar variance and helicity, and are defined in terms of the
viscosity $\nu$ and the thermal diffusion coefficient $\alpha$. The
kinetic energy and passive scalar variance have positive-definite
dissipation rates, while helicity does not.  The latter fact, however, does
not in itself preclude a helicity inertial-range
(\cite{CCE03,KurTayMat04b}). 
Equations (\ref{scalinglaws}) are three of the
few exact, nontrivial results known in the theory of statistical
hydrodynamics. They are valuable benchmarks for the study of high
Reynolds number ($Re$) turbulence in both experiments (\cite{DhrTsuSre97,MydlarskiL:Passsh,CHAMBERSAJ:ATMEOP}) and numerical simulations
(\cite{SVBSCC96, GotFukNak02, TayKurEyi03, KurTayMat04b}). 
\commentout{(BW: Maybe
say we're adding another equation to this list? SK: I like this idea.)}

In section \ref{KH} we derive the K\'arm\'an-Howarth equation for the
two-point correlation function of $q$ starting from the Boussinesq
equations for rotating and stratified flows.  As noted by
\cite{HerKerRot94} for zero rotation and stratification, the flux and viscous-diffusion
terms are not necessarily separated in scale (or wavenumber).  
 However, in statistically homogeneous flow with $Re
\rightarrow \infty$ and fixed Prandtl number $Pr$, there exists the possibility of a
range of scales wherein the viscous-diffusion rate is at least
sub-dominant, if not negligible, compared to the transfer
rate. Assuming such a range of scales exists at sufficiently high
$Re$, we recover a balance between the divergence of the third-order
correlation between velocity and potential vorticity, and the
production-dissipation rate of potential enstrophy.

In section \ref{Limits} we consider different limits of the Rossby 
($Ro$) and Froude ($Fr$) 
numbers.  
\commentout{ Since we are considering globally anisotropic systems,
we must be careful when we make the local isotropy assumption in order
to proceed along the lines of the K41 theory. Local isotropy implies
that two-point statistics are invariant under arbitrary rigid
rotations at sufficiently small scales, typically separated from
anisotropic boundaries and/or forced scales.} 
When rotation and
stratification are small ($Ro$ and $Fr$ are large), we can reasonably
hypothesize local isotropy at small-enough scales.  Then, an exact
scaling law analogous to
Eqs.~(\ref{scalinglaws}) arises for the third-order mixed
correlation of $\bm{ u}$ and $q$, with a prefactor of $2/3$. In the
QG limit ($Ro$ and $Fr$ are small), the dependence on $Ro$ and $Fr$
may be scaled out in a coordinate system with vertical stretching. For
QG scales in the stretched coordinates, we again assume local isotropy
and deduce the same linear scaling law for the mixed third-order
correlation.  In section \ref{Disc}, we tabulate several intermediate limits in $Ro$
and $Fr$; in some cases the correlation dynamics allow for a scale
separation between flux terms and viscous-diffusion terms, and hence
the clear possibility for an inertial range of scales dominated by
potential enstrophy flux.  \commentout{ Motivated by the strong result of
\cite{Charney71}, we speculate on the general role of $Q$ for 
constraining energy transfer (section \ref{Disc}).}

\section{The K\'arm\'an-Howarth equation for potential vorticity $q$}\label{KH}
The Boussinesq equations for rotating, stably stratified and incompressible flow are:
\begin{equation}
\frac{D}{Dt}\bm{u} + f\hat{\bm z}\times \bm{u} + 
\frac{1}{\rho_0}\nabla p + \frac{\tilde\rho}{\rho_0}g\hat{\bm{z}} = 
\nu\nabla^2 \bm{u}, \quad \grad\cdot \bm{u} = 0,\quad
\frac{D}{Dt}\tilde\rho - bw = \kappa \nabla^2 \tilde \rho,
\label{bous}
\end{equation}
where $D/Dt = \partial_t + (\bm{u} \cdot \grad)$,
$\bm{u}$ is the velocity, $w$ is its vertical component, 
$p$ is an effective pressure, 
$f=2\Omega$ is the Coriolis parameter, $\Omega$
is the constant background rotation rate, $\nu = \mu/\rho_0$ is the 
kinematic viscosity and $\kappa$ is the mass
diffusivity coefficient.
The total density is $\rho(\bm{x}) = \rho_0 - bz + \tilde\rho(\bm{x})$,
 where $\rho_0$ is the constant background, $b$ is also 
constant and larger than zero for stable stratification, 
and $\tilde \rho$ is the density fluctuation such that $| \tilde \rho | 
\ll | bz | \ll \rho_0$.

We begin with the equation for $q$ in
rotating, stably stratified flow (\cite{EM98}) with periodic boundary
conditions:
\begin{eqnarray}
\frac{D~q}{Dt}=\frac{D} {Dt}(\bm{\omega}_a \cdot {\nabla} \rho) &=&
\nu{\nabla^2} \bom \cdot {\nabla}\rho +
\kappa\nabla^2 {\nabla}\tilde\rho\cdot \bom_a
\label{vis-pv-bous}
\end{eqnarray}
where $\bom_a = \bom + f \hat{\bm{z}}$ is the absolute vorticity and $\bm{\omega} = {\nabla} \times \bm{u}$ is the relative vorticity.
The equation for the potential enstrophy $Q = q^2/2$ is obtained by
multiplying Eq.\ (\ref{vis-pv-bous}) by $q$. The
mean production-dissipation rate of $Q$ over the domain is then given by:
\begin{equation}
\frac{\partial}{\partial t} \langle Q \rangle = \nu\langle
{q~\nabla^2} \bm{\omega} 
\cdot {\nabla}\rho \rangle + \kappa\langle q~\nabla^2
{\nabla}\tilde\rho\cdot \bm{\omega}_a \rangle 
= - \varepsilon_Q
\label{epsQ-bous}
\end{equation} 
where $\langle\cdot\rangle$ denotes volume integration over the
periodic domain.
\commentout{
and we have used
\begin{equation}
\langle \frac{D} {Dt}(Q)\rangle =
\frac{\partial}{\partial t} \langle Q \rangle + \langle
\bm{u\cdot\nabla} Q\rangle = \frac{\partial}{\partial t} \langle Q
\rangle + \langle {\nabla\cdot u} Q\rangle =
\frac{\partial}{\partial t} \langle Q \rangle.
\end{equation}}
The quantity $\varepsilon_Q$ is not
sign-definite, allowing for $both$ production and dissipation,
of potential enstrophy (see also \cite{HerKerRot94}).

\subsection{Equation for the two-point correlation function of $q$ in homogeneous flow}
Following \cite{EM98}, one may define non-dimensional parameters:
\begin{equation}
\displaystyle
Ro = \frac{U}{Lf}, \quad Fr = \frac{U}{LN},\quad \Gamma = \frac{BgL}{U^2},\quad 
Pr = \frac{\nu}{\kappa},\quad \Lambda = \Gamma Fr^2, \quad Re = \frac{UL}{\nu}
\end{equation}
\ni where $U$, $L$, $L/U$ are the characteristic velocity,
length scale and time scale, and
$B\rho_0$ is the characteristic scale for the density fluctuations
($B$ is a dimensionless constant).  The buoyancy frequency $N$ is
given by $N = (g b/\rho_0)^{1/2}$.   For the limiting cases 
$Fr \rightarrow 0$ and $Fr \rightarrow \infty$, conservation of
total energy $(\bm{u}\cdot \bm{ u} + \tilde{\rho}^2)/2$ requires
that $\Gamma = 1/Fr$, and thus $\Lambda = Fr$.  The latter equality is
assumed throughout this work.  Then the
characteristic velocity is given by $U = Bg/N = B(\rho_0 g/b)^{1/2}$.

With the above definitions, the non-dimensional form of the
equation for $q$ as measured at point $\bm{x}$ becomes
\begin{eqnarray}
\frac{D}{Dt} [\bom \cdot \grad \tilde \rho + Ro^{-1} 
\frac{\partial \tilde \rho}{\partial z} - Fr^{-1} \omega_3 ] &=& 
Re^{-1}(Fr^{-1} \nabla^2 \omega_3  -  \grad \tilde \rho \cdot \nabla^2 \bom) \nonumber\\
&-& (RePr)^{-1} (Ro^{-1} \nabla^2 \frac{\partial \tilde \rho}{\partial z}
 + \bom \cdot\nabla^2  \grad \tilde \rho)
\label{pv_nondimensional}
\end{eqnarray}
\ni where all variables are non-dimensional and $\omega_3$ is the vertical component
of $\bom$.   From now on we include in $q$
only the contributions including the fluctuations $\bm{\omega}$ and $\tilde{\rho}$
\begin{equation}
q = \bom \cdot \grad \tilde \rho  + Ro^{-1} 
\frac{\partial \tilde \rho}{\partial z} - Fr^{-1} \omega_3,
\label{qnd}
\end{equation}
\ni
and we work entirely in the non-dimensional units.

One may write down an equation identical to (\ref{pv_nondimensional}) 
for $q'$, the potential vorticity at point 
$\bm{x}'$. Cross-multiplication and summing of the two resulting equations yields the equation for the two-point quantity $q q' = q(\bm{ x}) q(\bm{ x}')$:
\begin{eqnarray}
\frac{{\sf d}}{{\sf d}t}(q q') =\frac{{\sf d}}{{\sf d}t} &\Big [& (\omega_i \pai \tilde \rho)(\omega_j' \pajp 
\tilde\rho')
+ Ro^{-1} \{ \frac{\partial \tilde\rho}{\partial z}
\omega'_i \paip \tilde\rho' +
\frac{\partial \tilde\rho'}{\partial z'}
\omega_i \pai \tilde\rho \} - Fr^{-1} \{ \omega_3 \omega_i' \paip \tilde\rho' + 
\omega'_3 \omega_i \pai \tilde\rho \}\nonumber\\
&+&  Ro^{-2}  \frac{\partial \tilde\rho}{\partial z}
\frac{\partial \tilde\rho'}{\partial z'} 
- Ro^{-1}Fr^{-1} \{ \omega_3 \frac{\partial \tilde\rho'}
{\partial z'} + \omega_3' \frac{\partial \tilde\rho}
{\partial z}\} + Fr^{-2}\omega_3 \omega'_3 \Big]= 
\nonumber\\
Re^{-1}&\Big[& (\omega_k \pak \tilde\rho)(\pajp^2 \omega_i')(\paip \tilde\rho') + (\omega_k' \pakp \tilde\rho')
(\paj^2 \omega_i)(\pai \tilde\rho)
\nonumber\\
&+& Ro^{-1} \{
\frac{\partial \tilde\rho}{\partial z} (\pajp^2 \omega_i')
(\paip \tilde\rho') + 
\frac{\partial \tilde\rho'}{\partial z'} (\paj^2 \omega_i)
(\pai \tilde\rho)  \} 
\nonumber\\
&-& Fr^{-1} 
\{ \omega_3 ( \pajp^2 \omega_i') (\paip \tilde\rho')
+ \omega_3' ( \paj^2 \omega_i) (\pai \tilde\rho) +
(\omega_i \pai \tilde\rho)(\pajp^2 \omega_3')
+ (\omega_i' \paip \tilde\rho')(\paj^2 \omega_3) \}
\nonumber\\
&-& Ro^{-1} Fr^{-1}
\{ \frac{\partial \tilde\rho}{\partial z} \pajp^2 \omega_3' 
+ \frac{\partial \tilde\rho' }{\partial z' } \paj^2 \omega_3 \}
+Fr^{-2}\{ \omega_3 \pajp^2 \omega_3'+ \omega_3' \paj^2
\omega_3 \} \Big] + \label{qqprime_nondimensional}\\
Re^{-1}Pr^{-1}&\Big[& 
(\omega_k \pak \tilde\rho)\omega_i' (\paip \pajp^2 \tilde\rho')
+ (\omega_k' \pakp \tilde\rho')
\omega_i (\pai \paj^2 \tilde\rho) 
\nonumber\\
&+& Ro^{-1}\{
\frac{\partial \tilde\rho}{\partial z}\omega_i' (\paip
\pajp^2 \tilde\rho')
+ \frac{\partial \tilde\rho'}{\partial z'}\omega_i (\pai
\paj^2 \tilde\rho) +
(\omega_i \pai \tilde\rho)(\pajp^2 \frac{\partial \tilde\rho'}{\partial z'})
+ (\omega_i' \paip \tilde\rho')(\paj^2 
\frac{\partial \tilde\rho}{\partial z}) \}
\nonumber\\
&-& Fr^{-1}
\{ \omega_3 \omega'_i (\paip \pajp^2 \tilde\rho')
+ \omega_3' \omega_i (\pai \paj^2 \tilde\rho) \} 
\nonumber\\
&+& Ro^{-2}  
\{  \frac{\partial \tilde\rho}{\partial z} \pajp^2 
\frac{\partial \tilde\rho'}{\partial z'}
+ \frac{\partial \tilde\rho'}{\partial z'} \paj^2 
\frac{\partial \tilde\rho}{\partial z} \}
- Ro^{-1} Fr^{-1}\{ \omega_3 \pajp^2 \frac{\partial
  \tilde\rho'}{\partial z'} + \omega_3' \paj^2 
\frac{\partial \tilde\rho}{\partial z} \}
\Big]\nonumber
\end{eqnarray}
where ${\sf d}(q q')/{\sf d}t$ is defined by 
${\sf d}(q q')/{\sf d}t = \partial_t (q
q') + q' u_i \partial_i q + q u'_i\partial_{i'} q'$.
Here all primed variables denote their values at
$\bm{x}'$ and $\partial_{i'}$ denotes differentiation with respect to
$\bm{x_i}'$. 

Next one may express the equation in terms of the separation vector $\bm{r}$
between the points $\bm{x}$ and $\bm{x}'$, and rewrite the equation in terms of the new independent variables $\bm{r} = \bm{x}' - \bm{x}$ and $\bm {X} = (\bm{x} + \bm{x}')/2$ (see for example \cite{Hill02}), where
\begin{equation}
\partial_i  = -\partial_{r_i} +
\frac{1}{2}\partial_{X_i}; \;\; \partial_{i'}  = \partial_{r_i} +
\frac{1}{2}\partial_{X_i}; \;\;\partial_i^2 = \partial_i'^2 =
\partial_{r_i}^2 + \frac{1}{4}\partial_{X_i}^2.
\label{cov}
\end{equation}
The procedure to derive the equation for the two-point correlation
function is familar from, for example \cite{Frisch95,Hill02}. One can
perform a change of variables and then ensemble average the equation
assuming statistical homogeneity. Ensemble averaging commutes with the
derivative operations $\partial_{r_i}$ and
$\partial_{X_i}$. Statistical homogeneity, as in the case of periodic
boundary conditions with constant $N$ and $f$, implies that the derivative operation
$\partial_{X_i}$ acting on any statistic yields zero. The result is
\begin{eqnarray}
\frac{\partial~\langle q~q'\rangle }{\partial t} - 
\partial_{r_i}\langle q~q' (u_i - u'_i)\rangle  &=& 
Re^{-1}\partial_{r_i}\langle q \rho' \partial_{j'}^2 \omega'_{i} - q'  \rho \partial_j^2 \omega_{i} \rangle \nonumber\\
&+& (Re Pr)^{-1}\partial_{r_i}\langle q \omega'_{a_i} \partial_{j'}^2
\tilde\rho'- q'\omega_{a_i} \partial_j^2 \tilde\rho\rangle,
\label{qq'b-avg-hom}
\end{eqnarray}
where all factors of $Ro$ and $Fr$ are `hidden' in the
non-dimensional expressions for $q$ and $q^\prime$ defined by
Eq.~(\ref{qnd}), $\rho = -Fr^{-1}z + \tilde{\rho}$ and
$\bom_a = Ro^{-1} \hat{\bm{z}} + \bom$. Relation (\ref{qq'b-avg-hom})
is the non-dimensional K\'arm\'an-Howarth equation for the two-point,
second-order correlation function of $q$.  It describes the dynamics
of the two-point correlation of $q$ in a statistically homogeneous
Boussinesq flow, and is the starting point to consider various
limiting cases as described below.

\commentout{\begin{eqnarray}
\frac{\partial~(q~q') }{\partial t} &-& \partial_{r_i}(q~q' (u_i -
u'_i)) + \frac{1}{2}\partial_{X_i}(q~q' (u_i + u'_i)) =\nonumber\\
&&Re^{-1}\Big[(\partial_{r_i}(q \rho' \partial_{j'}^2 \omega'_{i})- q'
\rho \partial_j^2 \omega_{i} ) + \frac{1}{2}\partial_{X_i}(q' \rho
\partial_j^2 \omega_{i} + q \rho' \partial_{j'}^2 \omega'_{i}) \Big]
\nonumber\\ 
+ \; (Re Pr&)&^{-1}\Big[\partial_{r_i}( q \omega'_{a_i}
\partial_{j'}^2
\tilde\rho'- q'\omega_{a_i} \partial_j^2 \tilde\rho) +
\frac{1}{2}\partial_{X_i} (q'\omega_{a_i} \partial_j^2 \tilde\rho + q
\omega'_{a_i} \partial_{j'}^2 \tilde\rho')\Big]
\label{qq'b-cov}
\end{eqnarray}}


\subsubsection{Non-diffusive small-scale limit in homogeneous flow}
For finite $Re$ and $Pr$, the viscous-diffusion terms on the right-hand side of 
Eq.~(\ref{qq'b-avg-hom}) may not, in
general, act only at scales much smaller than the non-linear
(transfer) terms on the left-hand side.  
Furthermore, each individual term on the right-hand side is
not sign-definite, and thus allows for both production and dissipation
of $\langle q\;q^\prime \rangle$ at a given scale. The latter is reminiscent of the
sign-indefiniteness of the dissipation rate of helicity and helical
velocity statistics; despite this, it has recently been shown
that an inertial range of helicity can exist with an exact scaling law
recovered theoretically by \cite{Chkhetiani96} (see also
\cite{LPP97, GPP00, Kurien03}), as well as in simulations by
\cite{KurTayMat04b}.

We proceed with a particular order of limits, first $Re \rightarrow
\infty$ with $Pr$ fixed, followed by $r\rightarrow 0$. The
expectation is that this procedure will access a range of
(sufficiently small) scales such that the effect of the diffusion is
sub-leading compared to the transfer for a given scale. One may then
derive the leading-order balance between the flux and 
production-dissipation
terms. In the first limit, the right-hand side of
Eq.~(\ref{qq'b-avg-hom}) becomes negligible for given $r$. Then, in
the limit as $r
\rightarrow 0$, $\partial_t
\langle q q'\rangle \to 2 \partial_t
\langle Q\rangle = -2\varepsilon_Q$ and hence Eq.~(\ref{qq'b-avg-hom})
reduces to
\begin{eqnarray}
\nabla_{\bm r}\cdot \langle q~q' (\bm{u} - \bm{u}')\rangle = -2 \varepsilon_Q. 
\label{qq'b-avg-hom-nd}
\end{eqnarray}
Implicit in this step is the assumption that the production-dissipation rate
$\varepsilon_Q$ remains finite in the non-diffusive limit $Re \to
\infty$, $Pr$ fixed. This is analogous to the
\cite{K41} hypothesis that kinetic energy dissipation rate remains
finite in the inviscid limit. Similar assumptions were used for the
rates of dissipation of passive scalar variance and helicity to derive
Eqs.~(\ref{scalinglaws}). Equation ~(\ref{qq'b-avg-hom-nd}) is
analogous to the result for third-order velocity structure functions
in statistically homogeneous, high Reynolds number turbulence in the small scale limit (\cite{Frisch95}).

\section{Limiting cases in the rotation and stratification parameters}
\label{Limits}

Our starting point is the
K\'arm\'an-Howarth equation for homogeneous flows
Eq.~(\ref{qq'b-avg-hom}), upon which we impose
limits in $Ro$ and $Fr$ using Eq.~(\ref{qnd})
(see also Eq.~(\ref{qqprime_nondimensional})). 
In the large $Ro$, large $Fr$ limit, we derive an exact balance for 
the isotropic small scales between $\varepsilon_Q$ and the 
mixed third-order correlation of $\bm{u}$ and $q$.  
The scaling law thus derived, which we will call the
`$2/3$-law', is analogous to the established scaling laws for kinetic energy, 
passive scalar variance and helicity in Eqs.~(\ref{scalinglaws}). The same 2/3-law is shown to 
hold for the QG limit in a stretched 
coordinate system, presumably in a range of larger scales for 
which vertical stretching leads to local isotropy.

\subsection{Rotation and stratification are small (large $Ro$, large $Fr$)} 

The equations for large $Ro$ and large $Fr$ should tend toward the
equations for incompressible, variable density 3D flow. Consider the
following scalings:
\begin{equation}
\displaystyle
Ro = \frac{1}{\epsilon}\frac{N}{f},\;\;\; Fr = \frac{1}{\epsilon},\;\;\; Re \geq {\cal O}(1),\;\;\; Pr \;\; 
{\rm fixed}~\mbox{as}~\epsilon \rightarrow 0.
\end{equation}
After the change of variables according to (\ref{cov}), 
ensemble averaging and assuming homogeneity, the leading-order ${\cal O}(1)$ balance of
(\ref{qq'b-avg-hom}) is
\begin{eqnarray}
\frac{\partial~\langle q~q'\rangle }{\partial t} &-& 
\partial_{r_i}\langle q~q' (u_i - u'_i)\rangle = 
Re^{-1}[\partial_{r_i} \langle q \; \tilde\rho' \nabla'^2 
\omega'_i\rangle
-\partial_{r_i} \langle q' \tilde\rho \nabla^2 \omega_i\rangle]
\nonumber\\
&+& Re^{-1}Pr^{-1} [
\partial_{r_i} \langle q \; \omega_i' \nabla'^2\tilde\rho'
\rangle
-\partial_{r_i} \langle q' \omega_i \nabla^2\tilde\rho
\rangle ], \quad \mbox{where}~q \sim \bm{\omega}\cdot\grad \tilde \rho.
\label{qqprime_largefr_largero_r_ensemble_homo}
\end{eqnarray}
Note that, unlike in the equation for velocity correlations, 
the order of the derivative with respect to $\bm{r}$ is the same 
for the flux and viscous-diffusion terms.   
Thus we cannot, without further assumptions, 
expect an `inertial-range' transfer dominated by the 
flux of potential enstrophy in some range of scales. 
In this limit, the equations of motion are identical to those
investigated by \cite{HerKerRot94}, that is, the density is a passive
scalar and the momentum obeys the Navier-Stokes equations.
\commentout{\cite{HerKerRot94} performed a numerical study in Fourier space, and 
also noted the absence of a potential enstrophy inertial range 
at finite $Re$.}  
\commentout{
As already discussed, such an inertial range is
allowed in the limit $Re \rightarrow \infty$, and next we show
that it is characterized by a 2/3-law.}    

\subsubsection{Local isotropy}\label{homiso}
In the large $Ro$ and large $Fr$ limit, it is reasonable to further assume
local isotropy for sufficiently small scales, that is, invariance of the correlation
tensors under arbitrary rigid rotations.
Invariance with rotation by $\pi$ radians about each of the coordinate
axes yields the constraint that only the longitudinal components of the tensor correlations are non-zero. 
Then Eq.~(\ref{qqprime_largefr_largero_r_ensemble_homo}) reduces to
\begin{eqnarray}
\frac{\partial~\langle q~q'\rangle }{\partial t} - 
\nabla_{\bm{r}}\cdot \langle q~q' (\bm{u}_L &-& \bm{u}_L^\prime)\rangle
=
Re^{-1}{\nabla_{\bm{r}}}\cdot \langle q \rho' \partial_{j'}^2 \bm{\omega}_L^\prime - 
q'\rho \partial_j^2  \bm{\omega}_L \rangle  \nonumber\\
&+& (Re~Pr)^{-1} {\nabla_{\bm{r}}}\cdot \langle q \bm{\omega}^\prime_{aL}  \partial_{j'}^2
\tilde\rho'- q' \bm{\omega}_{aL} \partial_j^2 \tilde\rho\rangle,
\label{qq'b-avg-hom-iso}
\end{eqnarray}
where subscript $L$ denotes the longitudinal
component. Equation~(\ref{qq'b-avg-hom-iso}) 
is a K\'arma\'n-Howarth equation for two-point potential
vorticity statistics in 3D incompressible, variable density, statistically 
homogeneous and locally isotropic flow.
We may use the following isotropic forms for the scalar and
tensor correlations in Eq.~(\ref{qq'b-avg-hom-iso}):
\begin{eqnarray}
\langle q~q'\rangle = \langle q(\bm{x})q(\bm{x + r})\rangle = C(r)&; &~ \langle q(\bm{x})q (\bm{x + r})({u_i}(\bm{x}) - {u_i}(\bm{x+r}))\rangle = F(r)\frac{r_i}{r};\nonumber\\
\langle q \rho' \partial_{j'}^2{\omega}'_i - q'\rho \partial_j^2 {\omega_i} \rangle =G_1(r)\displaystyle\frac{r_i}{r}&;&
~\langle q {\omega}'_{a_i}  \partial_{j'}^2\tilde\rho'- q'\omega_{a_i} \partial_j^2 \tilde\rho\rangle = G_2(r)\displaystyle\frac{r_i}{r}
\end{eqnarray}
where $C(r)$, $F(r)$, $G_1(r)$ and $G_2(r)$ are scalar functions of $r$. Substituting these isotropic forms into Eq.~(\ref{qq'b-avg-hom-iso}) and using
the identities
\begin{eqnarray}
\frac{\partial r}{\partial r_i} = \frac{r_i}{r};\quad 
\frac{\partial}{\partial r_i} = \frac{\partial r}{\partial r_i}\frac{\partial}{\partial r} = \frac{r_i}{r}\frac{\partial}{\partial r}
\end{eqnarray}
we find
\begin{equation}
\frac{\partial~ C(r)}{\partial t} - \frac{1}{r^2}\frac{\partial}{\partial r}\Big(r^2 F(r)\Big)=Re^{-1} \frac{1}{r^2}\frac{\partial}{\partial r}\Big(r^2
G_1(r)\Big) + (Re~Pr)^{-1} \frac{1}{r^2}\frac{\partial}{\partial r}\Big(r^2
G_2(r)\Big).
\label{qq'b-avg-iso-scalar}
\end{equation}
Relation (\ref{qq'b-avg-iso-scalar}) is the 
the scalar form of the K\'arm\'an-Howarth
equation for the second-order moment $\langle q q^\prime\rangle $ 
in locally isotropic flow.

\subsubsection{Statistically steady state, non-diffusive and small-scale limit}\label{23}
The analogy to the Kolmogorov $4/5$-law for longitudinal structure
functions may be recovered in the non-diffusive limit using the
following steps.  First, assume a statistically steady state in time.
Then, in order to observe the limit where the viscosity and mass
diffusion contributions are small, take the limit $Re
\rightarrow \infty$ with $Pr$ constant, eliminating the right-hand side
of (\ref{qq'b-avg-iso-scalar}).  Follow this with $r \rightarrow 0$ to obtain:
\begin{eqnarray}
\frac{\partial}{\partial t}~ C(r)|_{r=0} + \frac{\partial}{\partial
  t}~ \Big(\frac{\partial C(r)}{\partial r}\Big|_{r=0}r\Big) + \dots +
\frac{1}{r^2}\frac{\partial}{\partial r}(r^2 F(r))\Big|_{r\rightarrow0} = 0.
\label{qq'b-avg-iso-exp}
\end{eqnarray}
The mean dissipation rate of the potential enstrophy is
the leading order term of the time derivative:
\begin{equation}
\frac{\partial}{\partial t}(C(0)) = \frac{\partial}{\partial t}\langle
q^2 \rangle = 2\frac{\partial}{\partial t}\langle Q \rangle = -2\varepsilon_Q
\label{limeq}
\end{equation}
where it is assumed that the higher-order contributions to the time derivative in
Eq.~(\ref{qq'b-avg-iso-scalar}) vanish as $r\rightarrow 0$ in the statistically steady state. This is analogous to the assumption made to derive the K\'arm\'an-Howarth equation for helicity 
(\cite{Kurien03}). 
Next  use (\ref{qq'b-avg-iso-exp})-(\ref{limeq}) in (\ref{qq'b-avg-iso-scalar}), 
multiply by $r^2$ throughout, and integrate with respect to $r$ to find
\commentout{ 
\begin{eqnarray}
-2 \varepsilon_Q +\frac{1}{r^2}\frac{\partial}{\partial r}(r^2 F(r))\Big|_{r\rightarrow0} = 0
\label{qq'b-avg-isolim2}.
\end{eqnarray}
Finally, multiplication by $r^2$ throughout and integration with respect
to $r$ gives:}
\begin{equation}
F(r) \sim -\frac{2}{3}\varepsilon_Q r, \quad r \rightarrow 0.
\label{qq'b-avg-int}
\end{equation}
The constant of integration is zero assuming that the scalar
function $F(r)$ remains regular as $r\rightarrow 0$. Alternatively, 
we can write the `2/3-law' for the third-order correlation of
$q$ and velocity as:
\begin{equation}
\langle q(\bm{x})q (\bm{x + r}) (\bm{u}(\bm{x}) - \bm{u}(\bm{x+r}))\cdot 
\hat{\bm{r}})\rangle = -\frac{2}{3}\varepsilon_Q r
\label{23law}
\end{equation}
where $\varepsilon_Q$ is obtained from Eq.~(\ref{epsQ-bous}) in the large $Ro$, large $Fr$ limit,
\begin{equation}
 \varepsilon_Q = - (\nu\langle {q~\nabla^2} \bm{\omega} \cdot {\nabla}\tilde\rho \rangle + \kappa\langle q~\nabla^2 {\nabla}\tilde\rho\cdot \bm{\omega} \rangle).
\label{epsQ-bous-2}
\end{equation} 
The equation (\ref{23law}) is the potential enstrophy counterpart to the 
scaling laws presented in Eq.~(\ref{scalinglaws}).
As for the helicity case, $\epsilon_Q$ is not sign definite. As we have discussed above, this fact in itself does not exclude the possibility of an inertial range.

\subsection{The quasi-geostrophic limit}
The quasi-geostrophic (QG) limit is attained in the high-rotation, large-stratification (low $Ro$, low $Fr$) limit corresponding to 
\begin{equation}\label{qg_limits}
Ro = \epsilon \frac{N}{f},\;\;\;  Fr = \epsilon \;\; ,\;\;\;  Re \geq {\cal O}(1),\;\;\;
Pr \;\; {\rm fixed} \;\;{\rm as } \;\; \epsilon \rightarrow 0. 
\end{equation}
After change of variables, ensemble averaging and using homogeneity, the
leading order ${\cal O}(1/\epsilon^2)$ terms from
(\ref{qqprime_nondimensional}) are
$$
\frac{\partial~\langle q~q'\rangle }{\partial t} - \partial_{r_i}\langle q~q' (u_i - u'_i)\rangle
= Re^{-1} \Big(2 \parj \langle \omega_3 \omega_3^\prime \rangle
 - \frac{f}{N}
\{ \parj \langle \frac{\partial \tilde\rho}{\partial z} \omega_3^\prime \rangle
+ \parj \langle \frac{\partial \tilde\rho^\prime }{\partial z^\prime } \omega_3 
\rangle \}\Big)$$
\begin{equation} 
 + Re^{-1}Pr^{-1}\Big(\frac{f^2}{N^2} \parj \langle \frac{\partial \tilde\rho}{\partial z} 
\frac{\partial \tilde\rho^\prime}{\partial z^\prime} 
\rangle-\frac{f}{N}  \{ \parj \langle \omega_3 \frac{\partial
  \tilde\rho^\prime}{\partial z^\prime} \rangle  + 
\parj \langle \omega_3^\prime \frac{\partial
  \tilde\rho}{\partial z} \rangle \}
\Big), \quad q \sim \omega_3 - \frac{f}{N}\frac{\partial \tilde\rho}{\partial z}.
\label{qqprime_qg_r_ensemble_homo}
\end{equation}
In this limit, the viscous-diffusion terms on the 
right-hand side of (\ref{qqprime_qg_r_ensemble_homo})
are all second-order derivatives with respect to $\bm{r}$.  Therefore they
can be considered localized at small scales, allowing an inertial range in
which viscous-diffusion contributions are separated in scale 
from transfer contributions.
To focus on QG scales rather than small scales, we scale out the
dependence on $f$ and $N$ by using a stretched $z$-coordinate, $z_* = (N/f)z$ (\cite{Charney71}). Then the scaled $q$ is given by $q = \omega_3 -\partial \tilde\rho/\partial
z_*$, and all $z$-derivatives $(f/N) \partial/\partial_z$ in
(\ref{qqprime_qg_r_ensemble_homo}) become $\partial/\partial z_*$.
Such vertical stretching removes anisotropy in the QG scales;
one may imagine pancake-shaped eddies, stretched to become spherical
eddies.  Following sections \ref{homiso}
and \ref{23}, one may next derive the K\'arm\'an-Howarth equation
analogous to Eq.~(\ref{qq'b-avg-iso-scalar}), and then a scaling law
for the third-order mixed correlation function in an isotropic range of scales, identical to
Eq.~(\ref{23law}).  However, the separation distance $r$ is measured
in the stretched coordinates, and the mean dissipation rate for QG is
given by $$\varepsilon_Q = -\Big[Re^{-1}\Big\langle
\frac{\partial \tilde\rho}{\partial z} \nabla_*^2 \omega_3\Big\rangle
 + (Re Pr)^{-1}\Big\langle 
\omega_3 \nabla_*^2\frac{\partial \tilde\rho}{\partial
  z}\Big\rangle\Big],\quad
\nabla_*^2 = \frac{\partial^2}{\partial x^2} + 
\frac{\partial^2}{\partial y^2} + \frac{N^2}{f^2}
\frac{\partial^2}{\partial z_*^2}.$$

\section{Summary and Discussion}\label{Disc}
Our results for various limits of $Ro$ and 
$Fr$ are summarized in Table \ref{thetable}.  
In cases (ii-iv), an inertial range of
scales dominated by flux of potential enstrophy may be realized
at finite $Re$ because the viscous-diffusion terms are confined
to small scales.   For QG flow with $Ro, Fr \rightarrow 0$ (case ii) 
and $Re \rightarrow \infty$, the 2/3-law describes locally isotropic 
flow in a coordinate system with vertical stretching.
\commentout{For QG dynamics (case ii), we have derived a $2/3$-law
describing the third-order mixed correlation between potential vorticity 
$q \sim \omega_3 - \partial \tilde{\rho}/\partial z_*$ 
and $\bm{u}$, valid in a coordinate system with vertical stretching
$z_* = (N/f)z$.  The QG $2/3$-law (\ref{23law}) 
may be relatively easy to measure in numerical simulations, 
since in this case one need only compute the evolution of the
scalar streamfunction $\Psi$, where 
$\bm{u} = \grad_h \times \hat{\bm{z}}_* \Psi$, $\tilde{\rho}= 
\partial \Psi/\partial z_*$ and $\grad_h = \hat{\bm{ x}} \; \partial_x
+ \hat{\bm{y}}\; \partial_y$.}
\begin{table}
\centering
\begin{tabular}{ccccl}
Case & $Ro$     &      $Fr$ &    $q$     & viscous-diffusion terms \\
\hline
i & $\displaystyle\frac{1}{\epsilon}\frac{N}{f}$ &
$\displaystyle\frac{1}{\epsilon}$ & $\bm{\omega}\cdot\grad \tilde
\rho$ & $Re^{-1}\partial_{r_i}\langle(q \tilde\rho'
\partial^2_{j^\prime} \omega'_i - q' \tilde\rho \partial^2_{j}
\omega_i)\rangle $
\\ 
& & & & $+ (Re Pr)^{-1}\partial_{r_i}\langle(q
\omega_i'\partial^2_{j^\prime}\tilde\rho'
- q' \omega_i \partial^2_{j}\tilde\rho)\rangle$ \\
%
%
ii & $\displaystyle \epsilon \frac{N}{f}$ & $\epsilon$ & 
$\displaystyle\omega_3 - \frac{f}{N}\frac{\partial
  \tilde\rho}{\partial z}$ &
$\displaystyle Re^{-1} \parj \langle 2  (\omega_3 \omega_3^\prime)  -
\frac{f}{N} ( \frac{\partial \tilde\rho}{\partial z} \omega_3^\prime 
+ \frac{\partial \tilde\rho^\prime }{\partial z^\prime }
\omega_3)\rangle$
\\
 & & & &$ \displaystyle + (RePr)^{-1}\parj \langle \frac{f^2}{N^2}  \frac{\partial \tilde\rho}{\partial z} 
\frac{\partial \tilde\rho^\prime}{\partial z^\prime} -\frac{f}{N} (  \omega_3 \frac{\partial
  \tilde\rho^\prime}{\partial z^\prime}  + 
 \omega_3^\prime \frac{\partial
  \tilde\rho}{\partial z}) \rangle $ \\
%
%
iii & $\epsilon$ & ${\cal O}(1)$ & $\displaystyle\frac{\partial \tilde \rho}{\partial z}$ & 
$2(RePr)^{-1}\partial^2_{r_j}\langle q q'\rangle$ \\
%
%
iv & ${\cal O}(1)$ & $\epsilon$ & $\omega_3$ & 
$2Re^{-1}\partial^2_{r_j}\langle q q'\rangle$  \\
%
%
v & $\displaystyle\frac{1}{\epsilon}$ & ${\cal O}(1)$ & 
$Fr^{-1} \omega_3 + \omega_i \partial_i \tilde{\rho}$ & 
$  Re^{-1}\partial_{r_i}\langle  (q\rho^\prime \partial^2_{j^\prime}
\omega_i^\prime - q^\prime \rho \partial^2_j \omega_i)\rangle$ \\
& & & & $+ (Re Pr)^{-1}\partial_{r_i}\langle (q\omega_i^\prime \partial^2_{j^\prime}
\tilde{\rho}^\prime - q^\prime \omega_i \partial^2_j \tilde{\rho}) \rangle$ \\
%
%
vi & ${\cal O}(1)$ & $\displaystyle \frac{1}{\epsilon}$ & 
$\displaystyle Ro^{-1} \frac{\partial \tilde{\rho}}{\partial z} + \omega_i \partial_i 
\tilde{\rho}$ &
$ Re^{-1}\partial_{r_i} \langle  (q\tilde{\rho}^\prime \partial^2_{j^\prime}
\omega_i^\prime - q^\prime \tilde{\rho} \partial^2_j \omega_i) \rangle$\\
& & & & $+ (Re Pr)^{-1}\partial_{r_i} \langle ( q\omega_{a_i}^\prime
\partial^2_{j^\prime}
\tilde{\rho}^\prime - q^\prime \omega_{a_i} \partial^2_j \tilde{\rho}) \rangle$ \\
\end{tabular}
\caption{The form of $q$ and the viscous-diffusion terms of 
(\ref{qq'b-avg-hom}) in various cases. \label{thetable}}
\end{table} 
In the remaining three cases (i, v and vi), the possibility of a
potential-enstrophy inertial range at finite $Re$ is not as clear because the flux
and viscous-diffusion contributions can in principle intermingle at
all scales. Nevertheless, the $2/3$-law Eq.~(\ref{23law}) is obtained for
the $Ro, Fr \rightarrow \infty$ case (i), again for locally isotropic flow 
in the $Re \rightarrow \infty$ limit.    In the derivation of
the 2/3-law for both cases (i) and (ii), there is a prescribed sequence
of limits:  $Re \rightarrow \infty$ with $Pr$ fixed, followed by the separation $r
\rightarrow 0$.
\commentout{under the assumption of local
isotropy, and by taking a prescribed sequence of limits: $Re
\rightarrow 0$ with $Pr$ fixed, followed by the separation $r
\rightarrow 0$.}  
In case (i),
the $2/3$-law is additional statistical information describing 
3D turbulent flow with a passive scalar $\tilde{\rho}$. 
\commentout{since rotation and stratification effects
are negligible.}  
A point of difference from Eq.\ (1.1b) is that the
$2/3$-law for $q$ contains information about the
geometry and structures of the flow; recall that $\bom$ evolves as a
line element and $\grad \tilde{\rho}$ evolves as a surface element,
and thus that $q$ evolves as a volume element (\cite{Ertel42}).
\commentout{(\cite{Ertel42,MullerP:Ertpvt, HAYNESPH:Oncai}).}

The 2/3-laws are derived assuming local isotropy for some range of
scales: ``sufficiently small'' scales for nearly isotropic flow with
$Ro, Fr \rightarrow \infty$; larger QG scales for which vertical
stretching removes anisotropy in the case of $Ro, Fr \rightarrow 0$.
It is important to note that even if local isotropy is not strictly
realized, there is always an {\it isotropic component} of the flow
that could obey the 2/3-law (see \cite{TayKurEyi03}).  We expect the
2/3-laws to become part of the theoretical foundation of rotating and
stratified flows, and a benchmark for high-Reynolds number physical
and numerical experiments, just as Eqs.~(\ref{scalinglaws}) provide
a foundation for isotropic turbulence research.

An eventual goal is to elucidate the connection between energy and
potential enstrophy in different parameter limits, as pioneered by
\cite{Charney71} for QG dynamics. 
It is
instructive to write the conservation laws for energy and potential enstrophy:
\begin{eqnarray}
%
   \frac{\partial }{\partial t} \langle | \mathbf{v} |^2 + \tilde
   \rho^2 \rangle = Re^{-1} \langle \nabla^2 |\mathbf{v} |^2 \rangle +
   (RePr)^{-1}\langle \nabla^2 \tilde \rho^2 \rangle,\nonumber\\
%
%
   \frac{\partial \langle Q \rangle}{\partial t}  = 
   Re^{-1}((Fr^{-1} \langle q \nabla^2 \omega_3\rangle - \langle
   q \grad \tilde \rho \cdot \nabla^2 \bom \rangle) 
   -Pr^{-1} (Ro^{-1} \langle q \nabla^2 \frac{\partial \tilde
     \rho}{\partial z}\rangle + \langle q~\bom \cdot\nabla^2 \grad
   \tilde \rho \rangle)),\nonumber
\end{eqnarray}

\ni
where $q = \bom \cdot \grad \tilde \rho + Ro^{-1} \partial \tilde \rho/
\partial z - Fr^{-1} \omega_3 $.  The
energy conservation law does not depend on either $Ro$ or $Fr$, while
the conservation law for potential enstrophy does (see also Table \ref{thetable}). 
Thus potential enstrophy may constrain energy
transfer among different scales in cases other than the QG limit, an
exciting possibility for future research.
\commentout{information that is clearly of consequence for
accurate parameterization of turbulence in numerical models of large-scale
geophysical flows.}

\acknowledgements
We thank D.\  Holm and F.\   Waleffe for helpful discussions. LS
is grateful for the support of NSF-DMS-0305479, and for sabbatical
funds provided by LANL through CCS-2, CNLS and IGPP.~SK and BW were
funded by DOE contract W-7405-ENG-36. The authors
acknowledge funding from NSF Collaborations in Mathematical
Geosciences (NSF-DMS-0529596), and from DOE ASCR program in Applied
Mathematics Research.

\bibliographystyle{jfm}
\bibliography{../../../Bibs/jfm-pv,../../../Bibs/heltime_pre,../../../Bibs/two15law_jfm,../../../Bibs/PV,../../../Bibs/TurbBousinesq}

\end{document}